\documentclass[
  reprint, %
  superscriptaddress, %
  amsmath,amssymb, %
  aps, %
  prb, %
  longbibliography %
]{revtex4-2}

\usepackage[english]{babel}
\usepackage{amsmath,amssymb}
\usepackage{graphicx}

\usepackage{dsfont}
\usepackage{comment}
\usepackage{physics}
\usepackage{mathtools}
\usepackage{booktabs}
\usepackage{array}
\usepackage{tikz}
\usepackage{xcolor}

\usetikzlibrary{matrix,arrows.meta,calc}

\newcommand{\bbZ}{\mathbb{Z}}

\newcommand{\id}{\mathds{1}}
\newcommand{\ii}{\mathrm{i}}
\newcommand{\Sig}{\operatorname{Sig}}
\newcommand{\Pf}{\operatorname{Pf}}
\newcommand{\sgn}{\operatorname{sgn}}

\newcommand{\Kc}{\mathcal{K}}

\usepackage[colorlinks=true, allcolors=blue]{hyperref}

\begin{document}

\title{Spectral localizers in phase space: \texorpdfstring{$\bbZ$}{Z} and \texorpdfstring{$\bbZ_2$}{Z2} invariants for \texorpdfstring{\\}{} higher-order, semimetal, and valley modes}

\author{Lucien Jezequel}
\affiliation{Department of Physics, KTH Royal Institute of Technology, 106 91, Stockholm, Sweden}
\author{Christophe Mora}
\affiliation{Universit\'e Paris Cit\'e, CNRS, Laboratoire Mat\'eriaux et Ph\'enom\`enes Quantiques, 75013 Paris, France}

\begin{abstract}
Topologically protected gapless modes emerge across multiple platforms, ranging from nodes in the band structure to boundary states in topologically non-trivial insulators, with each case governed by distinct topological classifications and invariants. Here, we generalize the spectral localizer method to establish a unified framework for topological gapless modes. The spectral localizer reduces the original Hamiltonian to a zero-dimensional operator by recursively applying quantum confining operators, which need not be diagonal in either position or momentum space, as previously assumed. In phase space, the zero-manifold of the semiclassical confining operators forms a shell enclosing the gapless mode of interest, analogous to the Teo--Kane framework for topological defects. The localizer yields a zero-dimensional invariant, such as a signature, a determinant sign, or a Pfaffian sign, capturing both $\bbZ$ and $\bbZ_2$ classifications. This single construction thus encompasses diverse scenarios, including hinge modes in higher-order topological insulators, individual Dirac nodes in semimetals, and valley-selective topological boundary modes. We illustrate our approach using lattice models in symmetry classes A, AIII, AII, and D, verifying that topological transitions coincide with a vanishing gap in the spectral localizer operator.

\end{abstract}

\maketitle

\section{Introduction}

Topological phases are usually characterized by invariants defined for gapped Hamiltonians. In the standard bulk-boundary picture, a change of bulk invariant forces low-energy modes to appear at an interface \cite{bernevig2013topological}. Protected modes also arise away from ordinary boundaries: as bulk nodes in topological semimetals \cite{Burkov_2016,Armitage_2018}, as valley-polarized modes in graphene \cite{Zhang_2013,Martin_2008,Rycerz_2007}, or as hinge and corner states in higher-order topological insulators (HOTIs) \cite{Benalcazar_2017,Schindler_2018}. These systems are often discussed with different tools, but they share a common feature: the protected modes are separated from the rest of the spectrum in the joint phase space $(x,k)$, either in position, in momentum, or in both.

This point of view is especially useful for finite, disordered, or non-periodic systems where momentum-space invariants such as Chern or winding numbers are ill defined. In that case one needs local diagnostics. Real-space Chern and winding markers provide one route, by rewriting momentum-space invariants in real space \cite{Bianco_2011,Prodan_2009,Huang_2018,Hannukainen2022}. The \textit{spectral localizer} provides another route \cite{LORING2015,Loring_2010,Loring_2017,Loring_2018,LozanoViesca_2019,SchulzBaldes_2024review}. It combines the Hamiltonian with localization operators into an auxiliary gapped operator, whose index detects the topology near a target region. For $\bbZ$ topological insulators the two routes are in fact equivalent \cite{jezequel2026equivalence}. For $\bbZ_2$ invariants, real-space markers have been constructed with the help of an auxiliary operator that commutes, at least approximately, with the projector onto the occupied states \cite{Hannukainen2022,kvorning2026}. The spectral localizer provides a complementary route, in which the $\bbZ_2$ index, the sign of a determinant or of a Pfaffian \cite{Loring_2017,Doll_2021}, follows directly from the anti-unitary symmetries of the Hamiltonian, without an auxiliary operator.

The aim of this work is to apply this idea beyond ordinary topological insulators. The modes of interest may be localized on hinges, at semimetal nodes in momentum space, or in mixed position-momentum regions as in systems exhibiting valley-selective topology~\cite{jezequel2023modeshell, jezequel2025modeshell}, in a way which can be applied to both $\bbZ$ and $\bbZ_2$ indices. The latter appears in systems protected by anti-unitary symmetries, such as time-reversal symmetry in quantum spin Hall systems \cite{KaneMele2005} and particle-hole symmetry in topological superconductors \cite{Sato_2017}, and also in crystalline settings \cite{Shiozaki_2017}. %

Local versions of the localizer have already been used for metals and gapless environments \cite{cerjan_local_2022,Dixon_2023,SchulzBaldes_2021,SchulzBaldes_2022}, for crystalline symmetries \cite{Cerjan_2024crystalline} and for periodic Hamiltonians through fuzzy tori \cite{Doll_2024}. More recently, a general spectral localizer method has been employed to generate real-space invariants for probing weak, strong, and crystalline topological phases~\cite{Chaou_2026}.
Our main contribution is a systematic phase-space formulation, showing that a single construction covers these diverse cases. By choosing confining operators with zeros forming a shell around the isolated gapless mode, the spectral localizer detects its topology. We apply this framework to higher-order hinge modes, Dirac nodes in topological semimetals, and valley gapless modes localized in both position and momentum.
The underlying principle is the mode-shell correspondence~\cite{jezequel2023modeshell,jezequel2025modeshell}, which relates the topology of gapless modes to that of a shell enclosing them in phase space.

The paper is organized as follows. Section~\ref{sec:spectral_localizer} presents the recursive construction, the phase-space shell, and its relation to the Teo--Kane classification of defects. Section~\ref{sec:hoti} applies it to the hinge modes of an extrinsic higher-order topological insulator, with and without disorder. Section~\ref{sec:semimetal} isolates single nodes in momentum space: the Dirac cones of graphene, the higher-order nodes of rhombohedral multilayer graphene, and a Dirac cone at a time-reversal-invariant momentum that carries both a $\bbZ$ and a $\bbZ_2$ index. Section~\ref{sec:valley} uses shells that mix position and momentum, for valley modes bound to smooth domain walls in graphene and in a Kitaev chain without hopping, whose global invariant vanishes while the valley-selective index does not.

\section{Dimensional reduction via the spectral localizer} \label{sec:spectral_localizer}

\subsection{Recursive dimensional reduction}

We consider a Hamiltonian $H$ defined in a $d$-dimensional space that hosts a topological gapless mode. The purpose of the spectral localizer method is to characterize the topology of this mode by implementing a dimensional reduction from $H$ to a fully gapped Hamiltonian $H'$ in zero dimension. The topological class and invariant of the gapless mode are then identified with those of $H'$, as determined by the tenfold classification \cite{AltlandZirnbauer1997,Kitaev2009,Ryu2010,Chiu_2016} in zero dimension shown in Table~\ref{tab:symmetry_clock}.

In zero dimension, three distinct topological invariants can be used depending on the symmetry class of $H'$.
The first invariant is the signature, namely the difference between the number of positive and negative eigenvalues of the Hamiltonian \cite{Loring_2018,LozanoViesca_2019}. It gives
\begin{equation}
    I(H)= \Sig(H')/2 \in \bbZ
    \label{eq:sigindex}
\end{equation}
which is non-trivial in classes A, AI, and AII.

If $H'$ is in class BDI, it can be represented by a real matrix with a chiral symmetry $CH'+H'C=0$. In the chiral basis it has the form
\begin{equation}
    H'=\begin{pmatrix}
        0&A\\A^{T} &0
    \end{pmatrix}
\end{equation}
where $A$ is a real invertible matrix. The sign of its determinant defines a topological invariant \cite{Loring_2017,Doll_2021},
\begin{equation}
    I(H)=\sgn (\det A)  \in \bbZ_2.
    \label{eq:detinvariant}
\end{equation}

Finally, if $H'$ is in class D, it can be written as a purely imaginary Hamiltonian $H'=\ii A$, where $A$ is real and antisymmetric. The corresponding invariant is the sign of the Pfaffian \cite{Loring_2017,Doll_2021},
\begin{equation}
    I(H)=\sgn (\Pf A)  \in \bbZ_2.
    \label{eq:pfaffinvariant}
\end{equation}

The dimensional reduction from $H$ to $H'$ can be achieved recursively by successively introducing a set of $\delta$ confining operators. The integer $\delta$ is the codimension, in phase space, of the shell that encloses the gapless mode (Sec.~\ref{sec:teokane}). For a mode localized in position, $\delta-1$ is simply its dimension~\cite{TeoKane_2010} (e.g., an edge state of a first- or higher-order topological insulator or a zero mode in $H$).
The successive Hamiltonians are labeled by $m$ and defined through the recursion:
\begin{align}
     H_{m-1}&=H_{m}+\kappa\, C G_{\delta-m+1}\label{eq:7a}\\
     H_{m-1}&=H_{m}\otimes \sigma_z+\kappa\, G_{\delta-m+1}\otimes\sigma_y \label{eq:7b}
\end{align}
where $H_{\delta}=H$ is the original Hamiltonian and $H_0 = H'$ the final spectral localizer in zero dimension. The two maps Eqs.~\eqref{eq:7a},~\eqref{eq:7b} alternate along the recursion: Eq.~\eqref{eq:7a} applies when $H_m$ exhibits a chiral symmetry $C$ while $H_{m-1}$ loses it, whereas Eq.~\eqref{eq:7b} applies when $H_m$ lacks chiral symmetry and $H_{m-1}$ acquires one with $C=\sigma_x$. Eq.~\eqref{eq:7b} additionally enlarges with $\sigma_z$ and $\sigma_y$ the size of the Hilbert space. A similar reduction scheme was implemented in Ref.~\cite{Chaou_2026} and weak indices can be extracted from intermediate $H_m$.

\begin{figure}[htbp]
    \centering
    \includegraphics[width=\linewidth]{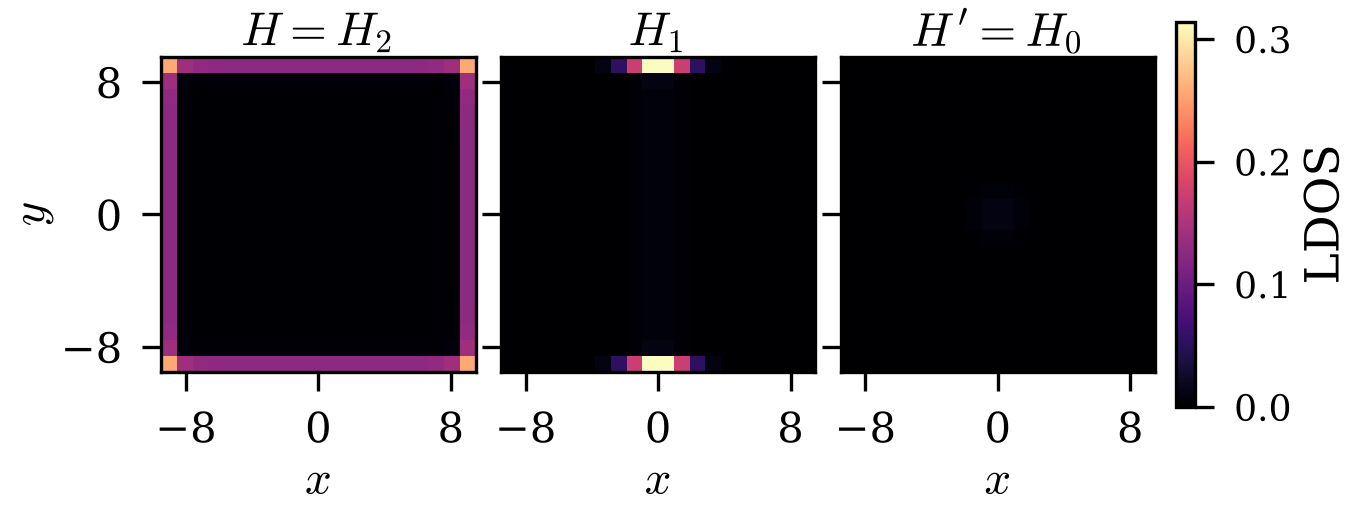}
    \caption{Local density of states (Appendix~\ref{ap:wigner}), in the $(x,y)$ plane, of the gapless modes of the Qi--Wu--Zhang Hamiltonian $H$ \cite{QiWuZhang2006}, of the partial localizer $H_1=H\otimes \sigma_z + \kappa X\otimes\sigma_y$ obtained after the first step \eqref{eq:7b}, and of the gapped localizer $H'=H_0=H_1+\kappa Y\otimes \sigma_x$ obtained after the second step \eqref{eq:7a} with $C=\sigma_x$: each confining operator removes one dimension from the gapless set, the one-dimensional edge modes being reduced to two zero-dimensional modes at $(0,\pm L/2)$, and then removed altogether. Parameters: $L=20$, $M=t=1$, $\kappa=0.3$, $\epsilon=0.42$.}
    \label{fig:dimensional-reduction}
\end{figure}

Remarkably, as discussed in Appendix~\ref{ap:Bottclock}, the map $H_{m}\rightarrow H_{m-1}$ follows the Bott clock of symmetry classes \cite{Chaou_2026} represented by arrows in Table~\ref{tab:symmetry_clock}.
Each recursion step progressively reduces the dimension of the gapless mode by one, and $\delta$ steps are required to fully gap the Hamiltonian. Consequently, the symmetry class of the model should be read from column $\delta$ in Table~\ref{tab:symmetry_clock} for the original Hamiltonian $H$, and from the zero-dimensional column for $H'$. We illustrate our recursive approach in Fig.~\ref{fig:dimensional-reduction} by applying it to a topological insulator (the Qi--Wu--Zhang model~\cite{QiWuZhang2006}, $H_{\vec k}=(M+t\cos k_x+t\cos k_y)\sigma_z+t\sin k_x\,\sigma_x+t\sin k_y\,\sigma_y$) with the localizer operators $G_1 = X$ and $G_2 = Y$. The low-energy densities shown in this and all other figures in the paper are obtained using a Gaussian spectral filter, as detailed in Appendix~\ref{ap:wigner}.

After completing the full recursion, the spectral localizer Hamiltonian $H' = H_0$ takes the form
\begin{align}
    &H'= H + \kappa\sum_{j=1}^{\delta} (C G_j)\, \Gamma_j\label{eq:8b} \\
    &H'= H\,\Gamma_0+\kappa\sum_{j=1}^{\delta} G_j \Gamma_j \label{eq:8a}
\end{align}
depending on whether $H$ has chiral symmetry (Eq.~\eqref{eq:8b}) or not (Eq.~\eqref{eq:8a}). $\Gamma_j$ are Clifford matrices obeying $\Gamma_i\Gamma_j+\Gamma_j\Gamma_i=2\delta_{ij}$ and built from the Pauli matrices introduced in Eqs.~\eqref{eq:7a},~\eqref{eq:7b}, with $\Gamma_0=\ii^{\,\delta(\delta-1)/2}\,\Gamma_1\cdots\Gamma_{\delta}$ for even $\delta$.
For a chiral $H$, the recursion starts with Eq.~\eqref{eq:7a} and gives
\begin{equation}
    H'=H\,\Gamma_0+\kappa\,C G_1\,\Gamma_0+\kappa\sum_{j=2}^{\delta}G_j\Gamma_j ,
\end{equation}
where $\Gamma_0=\sigma_z\otimes\cdots\otimes\sigma_z$, with one $\sigma_z$ per step~\eqref{eq:7b}. This is Eq.~\eqref{eq:8b} up to the change of basis
\begin{equation}
    U=P_+\otimes\id+P_-\otimes\Gamma_0 ,\qquad P_\pm=(\id\pm C)/2.
\end{equation}

\begin{table}[t]
\centering
\scriptsize
\renewcommand{\arraystretch}{1.05}

\begin{tikzpicture}[>=Stealth,scale=0.92,transform shape]

\tikzset{
  reduction/.style={
    ->,
    thick,
    shorten <=8pt,
    shorten >=8pt
  }
}

\matrix (m) [
    matrix of nodes,
    nodes={
        draw,
        anchor=center,
        align=center,
        text width=0.95cm,
        minimum height=0.5cm,
        inner sep=0pt,
        outer sep=0pt,
        font=\scriptsize,
        execute at begin node=\strut
    },
    column sep=-\pgflinewidth,
    row sep=-\pgflinewidth
]
{
Class & $T^2$ & $P^2$ & $C$ & $\delta=0$ & $\delta=1$ & $\delta=2$ & $\delta=3$ \\
A    & $0$  & $0$  & $0$  & $\mathbb{Z}$   & $0$            & $\mathbb{Z}$      & $0$ \\
AIII & $0$  & $0$  & $1$  & $0$            & $\mathbb{Z}$   & $0$               & $\mathbb{Z}$ \\
AI   & $+1$ & $0$  & $0$  & $\mathbb{Z}$   & $0$            & $0$               & $0$ \\
BDI  & $+1$ & $+1$ & $1$  & $\mathbb{Z}_2$ & $\mathbb{Z}$   & $0$               & $0$ \\
D    & $0$  & $+1$ & $0$  & $\mathbb{Z}_2$ & $\mathbb{Z}_2$ & $\mathbb{Z}$      & $0$ \\
DIII & $-1$ & $+1$ & $1$  & $0$            & $\mathbb{Z}_2$ & $\mathbb{Z}_2$    & $\mathbb{Z}$ \\
AII  & $-1$ & $0$  & $0$  & $2\mathbb{Z}$  & $0$            & $\mathbb{Z}_2$    & $\mathbb{Z}_2$ \\
CII  & $-1$ & $-1$ & $1$  & $0$            & $2\mathbb{Z}$  & $0$               & $\mathbb{Z}_2$ \\
C    & $0$  & $-1$ & $0$  & $0$            & $0$            & $2\mathbb{Z}$     & $0$ \\
CI   & $+1$ & $-1$ & $1$  & $0$            & $0$            & $0$               & $2\mathbb{Z}$ \\
};

\draw[reduction] (m-2-7.center) -- (m-3-6.center);

\draw[reduction] (m-3-6.center) -- (m-2-5.center);
\draw[reduction] (m-3-8.center) -- (m-2-7.center);

\draw[reduction] (m-5-6.center) -- (m-4-5.center);

\draw[reduction] (m-6-6.center) -- (m-5-5.center);
\draw[reduction] (m-6-7.center) -- (m-5-6.center);

\draw[reduction] (m-7-6.center) -- (m-6-5.center);
\draw[reduction] (m-7-7.center) -- (m-6-6.center);
\draw[reduction] (m-7-8.center) -- (m-6-7.center);

\draw[reduction] (m-8-7.center) -- (m-7-6.center);
\draw[reduction] (m-8-8.center) -- (m-7-7.center);

\draw[reduction] (m-9-6.center) -- (m-8-5.center);
\draw[reduction] (m-9-8.center) -- (m-8-7.center);

\draw[reduction] (m-10-7.center) -- (m-9-6.center);

\draw[reduction] (m-11-8.center) -- (m-10-7.center);

\end{tikzpicture}

\caption{Dimensional reduction induced by the spectral localizer on the Altland--Zirnbauer tenfold classification. The column $\delta$ is the number of confining operators needed to gap out the modes, that is, the codimension of the shell in phase space; boundary modes of a $\delta$-dimensional bulk have dimension $\delta-1$. Each arrow goes from an entry $(\text{class},\delta)$ to the entry $(\text{class}',\delta-1)$ reached after adding one real confining operator $G$, as derived in Appendix~\ref{ap:Bottclock}.}
\label{tab:symmetry_clock}
\end{table}

\subsection{Enclosing shell in phase space}

For $G_j=X_j$ one recovers the usual definition of the spectral localizer for topological insulators \cite{Loring_2010,cerjan_local_2022,cerjan_operator-based_2022,Cerjan_2024tutorial}. However, the recursive construction employed in this work extends beyond ordinary topological insulators and the confining operators $G_j$ can be chosen to be diagonal in position space, in momentum space, or as a combination or product of operators diagonal in both representations. Although our approach is purely quantum and we diagonalize a zero-dimensional Hamiltonian, it is useful for physical intuition to discuss the  semiclassical phase-space symbols $g_j(x,k)$ of the localizers $G_j$, defined through the
Wigner--Weyl transform. The surface or manifold where all $g_j(x,k)$ vanish, called the \emph{shell}, encloses the gapless mode of interest, in the spirit of the mode-shell correspondence~\cite{jezequel2023modeshell,jezequel2025modeshell}. The spectral localizer then detects its topology. A gapless set of dimension $\delta-1$ requires $\delta$ confining functions to be surrounded by the shell as in the Kl\'eman--Toulouse theory of topological defects~\cite{toulouse1976principles,Mermin_1979,TeoKane_2010}.

The condition for $H'$ to be gapped can be understood by taking its square:
\begin{align}
  &H'^2
  =
  H^2
  +
  \kappa^2\sum_j G_j^2
  \notag\\
  &
  +\kappa\sum_j[H,G_j]\,\Gamma_0\Gamma_j
  +\kappa^2\sum_{i<j}
  [G_i,G_j]\,\Gamma_i\Gamma_j ,
  \label{eq:multi-square}
\end{align}
where $\Gamma_0$ is replaced by the chiral operator $C$ in the chiral case \eqref{eq:8b}, and all sums run over $1\leq i,j\leq \delta$. This expression separates into two parts. The first one,
\begin{equation}
    Q=H^2
  +
  \kappa^2\sum_{j=1}^{\delta}G_j^2
\end{equation}
is a positive operator. The remaining term,
\begin{equation}\label{eq:R}
  \mathcal{R}= \kappa\sum_{j=1}^{\delta}[H,G_j]\,\Gamma_0\Gamma_j
  +
  \kappa^2\sum_{i<j}
  [G_i,G_j]\,\Gamma_i\Gamma_j ,
\end{equation}
with no definite sign, such that $H'^2=Q+\mathcal{R}$. A sufficient condition for a gapped $H'$ is that the smallest eigenvalue $\Delta_Q$ of $Q$ remains larger than the norm of $\mathcal{R}$, or~\cite{LORING2015,Cerjan_2024tutorial}
\begin{equation}
    \Delta_Q \;>\; \kappa\sum_{j}\bigl\|[H,G_j]\bigr\|
    +\kappa^2\sum_{i<j}\bigl\|[G_i,G_j]\bigr\| ,
    \label{eq:gapcriterion}
\end{equation}
This strict condition is often not necessary but establishes a convenient criterion for an estimate on the value of $\kappa$. Increasing $\kappa$ stabilizes the gap of $Q$ where $H$ itself is gapless but also increases $\mathcal{R}$ and a compromise is needed to entirely gap $H'$. The presence of commutators in Eq.~\eqref{eq:R} is crucial to ensure that $\mathcal{R}$ is bounded. They remain small as long as the confining operators or the Hamiltonian are relatively local in position and momentum space.

\subsection{Relation to the Teo--Kane classification of defects}
\label{sec:teokane}

The counting of confining operators in Eqs.~\eqref{eq:8b},\eqref{eq:8a}, and the column of Table~\ref{tab:symmetry_clock} from which the reduction starts, have a natural interpretation in the classification of topological defects by Teo and Kane~\cite{TeoKane_2010}, whose notation we follow. A defect in a $d$-dimensional insulator is enclosed by a sphere $S^D$ in position space, and away from the defect the system is described by a semi-classical Hamiltonian $\mathcal{H}(\vec k,\vec s)$ that varies slowly with the position $\vec s\in S^D$. $\vec k$ covers the d-dimensional torus $T^d$ and the topology of the defect is therefore that of a gapped Hamiltonian on the manifold $T^d\times S^D$. Teo and Kane showed that it is classified as a bulk insulator of dimension $\delta=d-D$, in the same symmetry class as the crystal: only the difference between the number of momentum and position directions matters.
The defect itself has dimension $d-D-1=\delta-1$, so that the gapless modes it binds are classified by their own dimension plus one, and not by the dimension of the system. The edge of a two-dimensional insulator ($d=2$, $D=0$) and the hinge of a three-dimensional higher-order insulator ($d=3$, $D=1$) both have $\delta=2$ and fall in the same column in Table~\ref{tab:symmetry_clock}.

Our shell construction is analogous to the Teo--Kane manifold.
Within the $2d$-dimensional phase space, the subspace $T^d\times S^D$ is defined by $2d-(d+D)=\delta$ independent constraints, which determines its codimension. In our spectral localizer method, the shell is likewise defined by $\delta$ semiclassical conditions $g_j(x,k)=0$.
The number of confining operators is thus given by the Teo--Kane parameter $\delta$, which  determines the topological classification of Table~\ref{tab:symmetry_clock}
as in Ref.~\cite{TeoKane_2010}. In fact, our recursive method explicitly identifies $\delta$ as the column index that fixes the topological classification as it is also the number of steps required to transform $H$ into $H'$ along the Bott clock.

Two features go beyond the original setting. First, Teo and Kane consider defects localized in position   and enclose them by a sphere in position space times the whole Brillouin zone. The shell considered in our work is not restricted to this form:
for a semimetal node, it is a sphere in momentum space at fixed position, while for the topological valley modes discussed in Sec.~\ref{sec:valley}, it mixes position and momentum. The interpretation of $\delta$ as a codimension in phase space remains unchanged and encompasses, in particular, the classification of point nodes~\cite{Chiu_2016}. Second, the Teo--Kane invariant is defined in terms of a slowly varying Bloch Hamiltonian, whereas the localizer requires only the operators $H$ and $G_j$ on a finite lattice. Consequently, the localizer index can be applied to disordered and inhomogeneous systems, for which $\mathcal{H}(\vec k,\vec s)$ is not available.

\section{Higher-order topological insulators}
\label{sec:hoti}

\begin{figure*}[t]
    \centering
    \includegraphics[height=5.2cm]{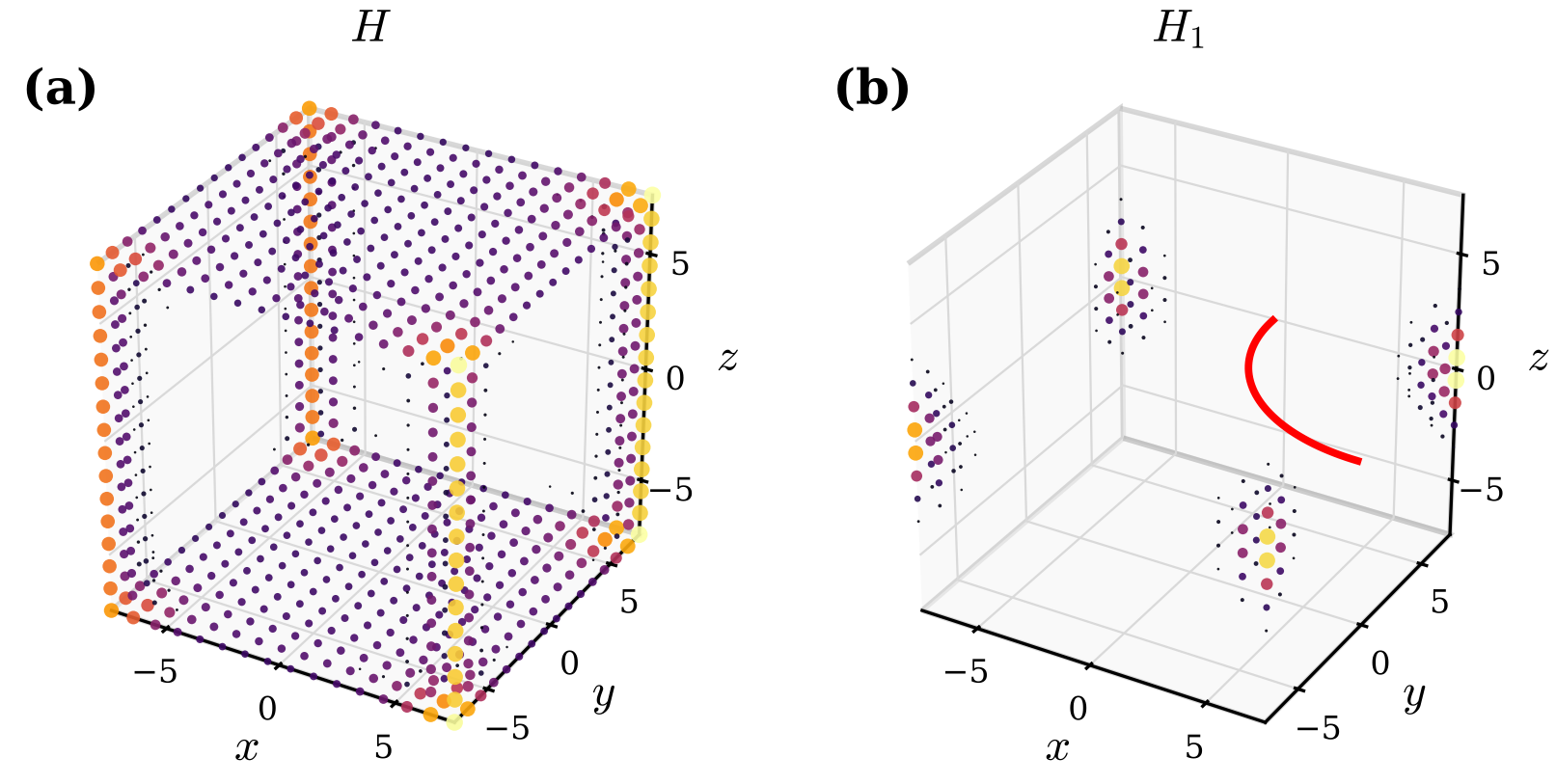}\hfill
    \includegraphics[height=5.2cm]{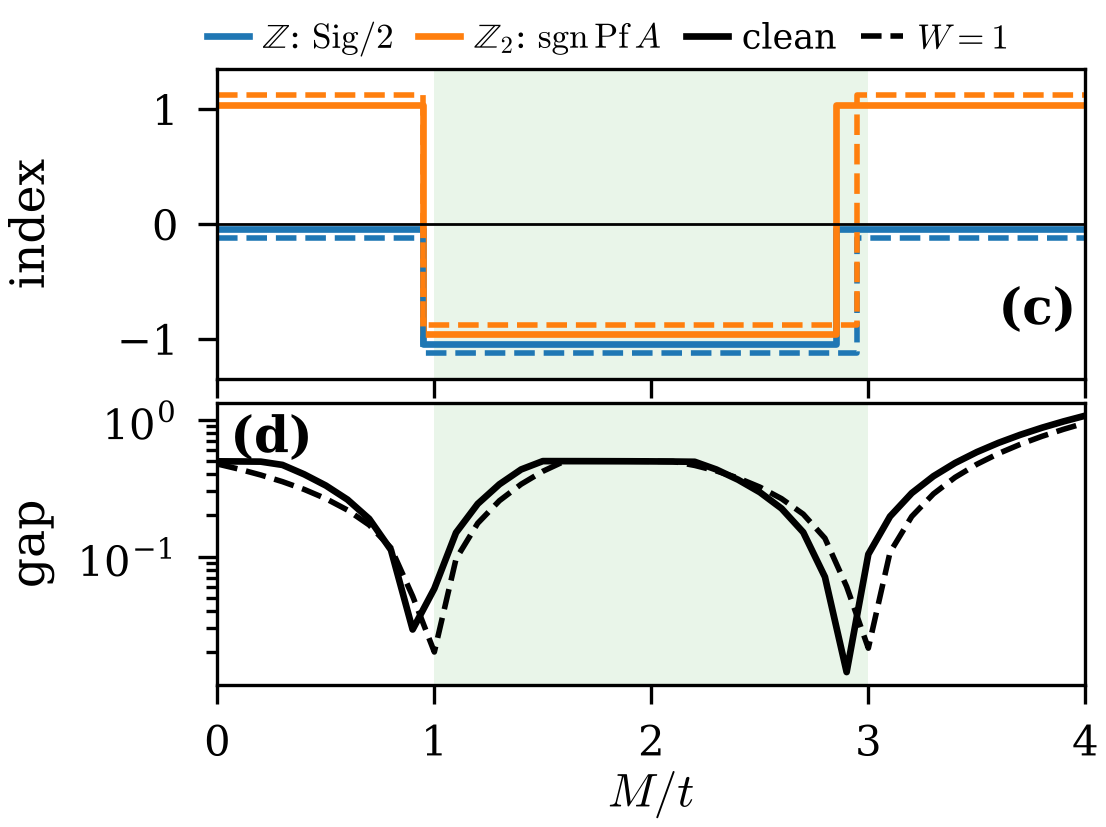}
    \caption{(a,b) LDOS (Appendix~\ref{ap:wigner}) of the low-energy states of $H$ and of the partial localizer $H_1$, for $L=16$ and $(M/t,\Delta,\lambda,\kappa)=(2,1,1,0.3)$; the shell $G_1=G_2=0$ is shown in red. (c) $\bbZ$ and $\bbZ_2$ indices of the models \eqref{eq:3DHOTI} and \eqref{eq:3DHOTIZ2} versus $M/t$, and (d) the localizer gap, common to both, for $L=20$ and $(\Delta,\lambda,\kappa)=(1,1,0.1)$. Dashed lines: one realization of on-site disorder uniform in $[-W,W]$ with $W=1$, half the clean bulk gap at $M/t=2$. The indices change only where the gap closes, close to the bulk transitions $M/t=1$ and $3$ (shaded: higher-order phase), with a shift that decreases as $1/L$. Curves on the same integer are offset for visibility.}
    \label{fig:HOTI}
\end{figure*}

Higher-order topological insulators can be classified as either intrinsic or extrinsic~\cite{Geier_2018,Trifunovic_2019}. In the extrinsic case, both the bulk and most of the boundaries are required to remain gapped, while gapless modes emerge at points or hinges and exhibit nontrivial topology. In contrast, intrinsic higher-order topology requires only a bulk gap for its protection, but is necessarily accompanied by crystalline symmetries.
Localizer invariants for such crystalline phases are constructed in Refs.~\cite{Cerjan_2024crystalline,Chaou_2026}. We discuss here only extrinsic topology and consider gapless modes localized on hinges. The shell surrounding them is given by the confining operators
\begin{equation}
    G_1=Z-z_0, \qquad
    G_2=\frac{(X-x_0)^2+(Y-y_0)^2-R^2}{2R} ,
    \label{eq:hoticonfining}
\end{equation}
where the hinge runs along the z direction at $(x_0,y_0)$. Classically, $G_1$ and $G_2$ localize on a circle with radius $R$ enclosing the hinge at $z=z_0$.

We consider the 3D model Hamiltonian, written here in momentum space,
\begin{equation}
\begin{split}
& H_\mathbf{k}
=
\left(M+t\sum_{i=x,y,z}\cos k_i\right)\tau_z\sigma_0
\\
&+\lambda\sum_{i=x,y,z}\sin k_i\,\tau_x\sigma_i
+\Delta(\cos k_x-\cos k_y)\tau_y\sigma_0 ,
\label{eq:3DHOTI}
\end{split}
    \end{equation}
which may host gapless hinge modes. Here, $\tau_i$ and $\sigma_i$ are Pauli matrices acting on two independent internal degrees of freedom. Originally introduced in Ref.~\cite{Schindler_2018}, this model provides an example of an intrinsic second-order topological insulator. The hinge modes are protected by crystalline symmetries, either by $C_4T$ symmetry or by a set of mirror symmetries. Instead, in the perspective of extrinsic topology, Eq.~\eqref{eq:3DHOTI} exhibits no symmetry and belongs to class A. Following the recursive spectral localizer method introduced in Sec.~\ref{sec:spectral_localizer}, $H$ is successively transformed by $G_1$ and $G_2$, taking $\delta$ from $2$ to $0$ along the arrows in Table~\ref{tab:symmetry_clock}, i.e., $\mathrm{A}\rightarrow\mathrm{AIII}\rightarrow\mathrm{A}$. The model is thus characterized by a $\mathbb{Z}$ topological index, which is nonzero for $1<|M/t|<3$, where a chiral hinge mode emerges, as illustrated in Fig.~\ref{fig:HOTI}.

A different symmetry class can be investigated by simply doubling the model. The Hamiltonian
\begin{equation}
    \tilde{H}_\mathbf{k}=\begin{pmatrix}
        H_\mathbf{k}&0\\0& H^*_{-\mathbf{k}}
    \end{pmatrix} \label{eq:3DHOTIZ2}
\end{equation}
satisfies time-reversal symmetry with $T=\mu_y\Kc$. $\mu_y$ is a Pauli matrix acting in the additional orbital space and $\Kc$ denotes complex conjugation such that $T^2=-\id$. The model therefore belongs to class AII and follows the sequence of symmetry classes
\begin{equation}
    \mathrm{AII}\;\longrightarrow\;\mathrm{DIII}\;\longrightarrow\;\mathrm{D},
\end{equation}
under the recursive spectral localizer method, as shown in Table~\ref{tab:symmetry_clock} (see also Appendix~\ref{ap:Bottclock}), predicting a $\bbZ_2$ topological invariant. As $H'$ sits in class D, it can be written in a basis where particle-hole symmetry acts as complex conjugation. Then $H'=\ii A$ with $A$ real and antisymmetric, and the $\bbZ_2$ index of the enclosed hinge modes is $I=\sgn\Pf(A)$. This is the localizer counterpart of the Pfaffian formulas obtained for real index pairings in Refs.~\cite{Loring_2017,Doll_2021}. The non-trivial topology is also obtained for $1<|M/t|<3$ and corresponds to the emergence of helical hinge states. Figure~\ref{fig:HOTI}(c) shows the Pfaffian index alongside the $\bbZ$ signature index of the chiral model Eq.~\eqref{eq:3DHOTI}.

It is worth emphasizing that the spectral localizer method does not rely on translation invariance. It is therefore robust to the addition of disorder, provided that the spectral gap of $H'$ remains open. We introduce an on-site potential drawn uniformly from $[-W,W]$ and diagonal in the orbital indices. For the helical model, the added disorder naturally preserves time-reversal symmetry. We use a single disorder realization with $W=t$, half the bulk gap of the clean model at $M/t=2$, and keep it fixed along the sweep. The resulting numerical topological indices for the chiral and helical models are shown as a function of the bulk parameter $M/t$ in Fig.~\ref{fig:HOTI}(c,d). Both indices keep their clean values deep in each phase and change only where the localizer gap closes.

\section{Topological semimetals}
\label{sec:semimetal}

So far, our analysis has been restricted to confining operators $G_j$ that are diagonal in position space in order to isolate edge modes in conventional (Sec.~\ref{sec:spectral_localizer}) and higher-order (Sec.~\ref{sec:hoti}) topological insulators. The spectral localizer method can also be extended to characterize the topology of isolated gapless modes in semimetals~\cite{SchulzBaldes_2021,SchulzBaldes_2022}. For a zero-mode in a model in $d$ dimension, $\delta = d+1$ confining operators are required for the spectral localizer: $d$ operators to confine in real space and one additional operator to encircle the zero in momentum space.

Consistently with the classification of point nodes~\cite{Matsuura_2013,Chiu_2016}, the topological index of such a gapless mode is read from Table~\ref{tab:symmetry_clock} in the column $\delta = d+1$, since a gapless mode in $d$ dimensions can be viewed as the boundary of a $(d+1)$-dimensional topological insulator in the same symmetry class.

\begin{figure*}[t]
    \centering
    \includegraphics[width=\textwidth]{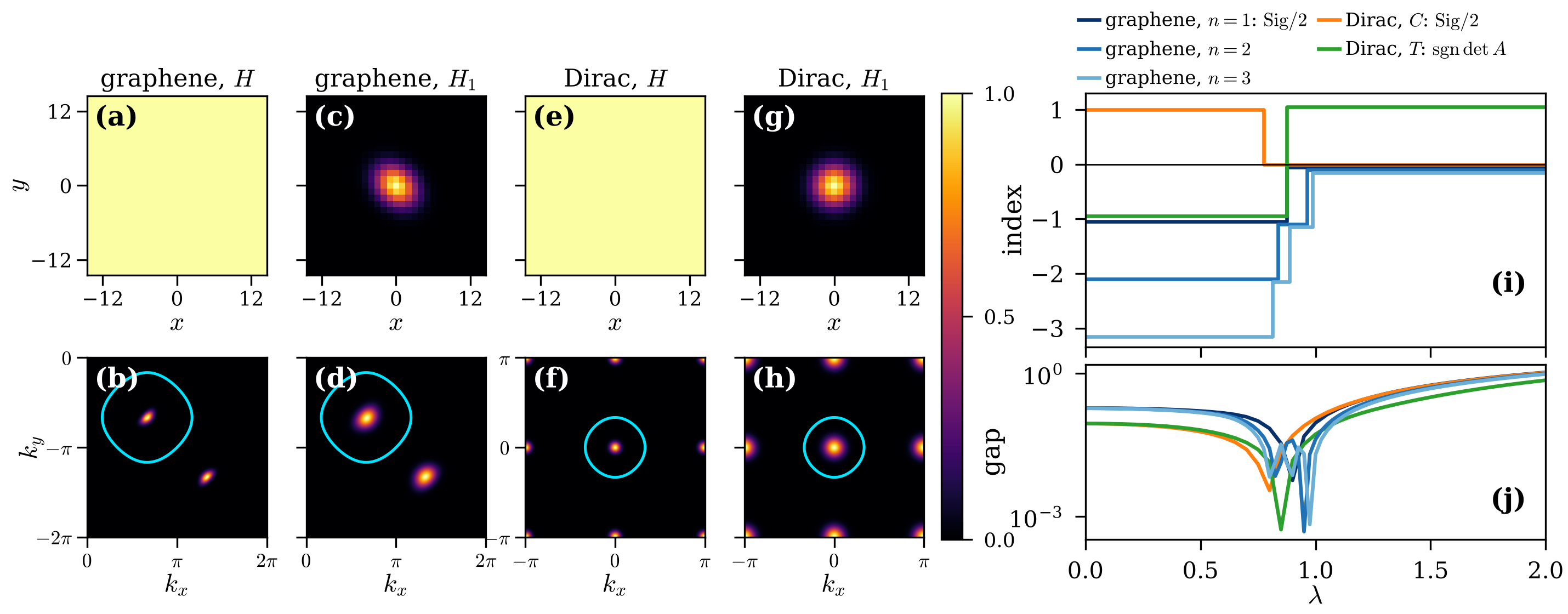}
    \caption{(a--h) Low-energy weight in position (top) and momentum (bottom) for $H$ and for the partial localizer $H_1$ (after $G_1=X$ and $G_2=Y$): graphene \eqref{eq:semimetaltopo} in (a--d), minimal Dirac model \eqref{eq:minimaldirac} in (e--h); periodic $29\times29$ lattice, $\epsilon=0.17$ (Appendix~\ref{ap:wigner}). Cyan: the shell $g_3=0$, which encloses one valley of graphene and the cone at the origin of the Dirac model. (i) Indices under the deformations $H-\lambda\sigma_x$ (half-signature: graphene, $n$-layer graphene with $\gamma_1=0.12\,t$, Dirac model) and $H+\lambda\,\id_2$ (determinant sign: Dirac model), and (j) the localizer gaps. Each index changes where its gap closes, close to the symbol predictions $\lambda=t$ for graphene and $\lambda=\sin(\pi/3)$ for the Dirac model. Curves on the same integer are offset for visibility. Parameters: $15\times15$ lattice, $\kappa=0.1$.}
    \label{fig:localizersemimetal}
\end{figure*}

\subsection{Graphene: a $\bbZ$ index from the sublattice symmetry}

We begin with the straightforward example of graphene, described as electron hopping on a honeycomb lattice.
Expressed in oblique coordinates, where $\sigma_\pm$ act on the sublattice degree of freedom, the system is described by
\begin{equation}
    \begin{split}
        H &= -t\bigl(\id+T_x^\dagger+T_y^\dagger\bigr)\otimes\sigma_+ +\mathrm{h.c.}\\
      H_{\vec k} &= \mathrm{Re}\,f(\vec k)\,\sigma_x-\mathrm{Im}\,f(\vec k)\,\sigma_y,
    \\[2pt]
      f(\vec k)&=-t\bigl(1+\mathrm{e}^{\ii k_x}+\mathrm{e}^{\ii k_y}\bigr),
    \label{eq:semimetaltopo}
    \end{split}
\end{equation}
where the second line is the Bloch Hamiltonian, which depends on the quasimomentum $\vec k$. The $T_{x_i}$ are discrete translation operators along the corresponding real-space directions. The function $f(\vec k)$ vanishes at exactly \emph{two} points of the Brillouin zone,
\begin{equation}
K=(2\pi/3,-2\pi/3), \qquad K'=(-2\pi/3,2\pi/3),
\label{eq:valleys}
\end{equation}
defining the two valleys of graphene. Gapless Dirac cones emerge from these two points,
and the corresponding Bloch states in their vicinity are delocalized in position space along both the $x$ and $y$ directions. Therefore, although these modes are diagonal (and thus localized) in momentum space, they must still be localized spatially, requiring three confining operators for the spectral localizer. The first two are purely spatial: $G_1=X$ and $G_2=Y$ localize near the origin. The third localizer selects one particular valley in momentum space, for instance
\begin{equation}
G_3=2-\bigl(\mathrm{e}^{-\ii K_x}T_x^\dagger+\mathrm{e}^{-\ii K_y}T_y^\dagger+\mathrm{h.c.}\bigr),
\end{equation}
isolates the Dirac cone in the $K$ valley. In momentum space, $G_3(\vec k)=2-2\cos(k_x-K_x)-2\cos(k_y-K_y)$ equals $-2$ at $K$ and $+4$ at $K'$ and at $\Gamma$, and its zeros form a closed curve encircling only $K$ [Fig.~\ref{fig:localizersemimetal}(b,d)]. Panels (a--h) of this figure show the low-energy weight in position and in momentum, two marginals of the same phase-space density (see Appendix~\ref{ap:wigner} for details).

Graphene exhibits chiral symmetry with $C=\sigma_z$, as well as time-reversal symmetry (TRS) mapping $K$ onto $K'$ and vice versa, corresponding to class BDI. However, isolating a single valley $K$ effectively breaks TRS, leaving only chiral symmetry and reducing  to class AIII. The dimensional reduction from $\delta=3$ then reads
\begin{equation}
    \mathrm{AIII}\rightarrow\mathrm{A}\rightarrow\mathrm{AIII}\rightarrow\mathrm{A},
\end{equation}
yielding a $\bbZ$ topological invariant $I_G$. We find $I_G = -1$ at $K$ and $I_G=+1$ at $K'$, that is, minus the winding number of $f(\vec k)$ around each cone.

We monitor the position and topology of the Dirac cone by adding the chiral-symmetry-preserving term $-\lambda \sigma_x$, with $\lambda\geq0$, to the graphene model Eq.~\eqref{eq:semimetaltopo}. This shifts the two Dirac cones~\cite{Montambaux_2009} along the line $k_x = - k_y$. As illustrated in Fig.~\ref{fig:localizersemimetal}, the Dirac cone exits the region delimited by $G_3$ for sufficiently large $\lambda$ and the topological index consequently drops to zero. Since the shell passes through the point $M=(\pi,-\pi)$ where the two cones annihilate at $\lambda=t$, the selected cone leaves the shell exactly when it annihilates with the cone of the other valley. With a smaller shell, the index would drop earlier, when the cone crosses the shell, as for the Dirac model of Sec.~\ref{sec:dirac}. In Fig.~\ref{fig:localizersemimetal}(i), the drop occurs at $\lambda\simeq0.89\,t$ rather than at $t$: the confinement by $G_1$ and $G_2$ broadens the cone in momentum [Fig.~\ref{fig:localizersemimetal}(d)], so that the localizer gap closes slightly before the center of the cone reaches the shell. On the same $15\times15$ lattice, the transition moves towards $t$ as $\kappa$ decreases, from $\lambda\simeq0.86\,t$ at $\kappa=0.2$ to $0.91\,t$ at $\kappa=0.05$.

\subsection{Rhombohedral multilayer graphene: nodes of higher order}

The same construction applies to rhombohedral (ABC-stacked) multilayer graphene, made of $n$ graphene layers in which the sublattice $B$ of each layer is coupled to the sublattice $A$ of the next one by a hopping $\gamma_1$. The details of the model are provided in Appendix~\ref{ap:multilayer}. We take $\gamma_1=0.12\,t$, close to its value in graphite. This coupling preserves the chiral symmetry, and at each valley the two bands closest to zero energy touch with a dispersion $|E|\propto|\vec q|^n$, $\vec q$ being measured from the K or K' point. The corresponding wavefunctions exhibit a winding of order $n$, equal to the number of layers~\cite{McCann_2006,Min_2008,Koshino_2009}. With the same operators $G_1=X$, $G_2=Y$ and $G_3$ as for the monolayer, the half-signature at $\lambda=0$ is $-n$, that is, $n$ times the monolayer value; we checked this for $n=1,\dots,4$. The deformation $-\lambda\sigma_x$, applied to every layer, moves the node of order $n$ along the same path as the monolayer cone, and the index drops from $-n$ to $0$ near the monolayer transition [Fig.~\ref{fig:localizersemimetal}(i,j)], in $n$ steps spread over a range of order $\gamma_1$, as detailed in Appendix~\ref{ap:multilayer}.

\subsection{Dirac cone with time reversal}
\label{sec:dirac}

The Dirac points in graphene occur at non-zero momenta and are therefore not invariant under time reversal. A closely related model can be constructed in which the Dirac points are shifted to time-reversal-invariant momenta (TRIM). This is achieved by the minimal two-band spin-orbit model
\begin{equation}
    H = \sin k_x\,\sigma_x+\sin k_y\,\sigma_y
    = \frac{T_x^\dagger-T_x}{2\ii}\,\sigma_x+\frac{T_y^\dagger-T_y}{2\ii}\,\sigma_y ,
    \label{eq:minimaldirac}
\end{equation}
in which $\sigma$ acts on a spin rather than on a sublattice, and time reversal is $T=\sigma_y\Kc$ with $T^2=-\id$. This model exhibits four Dirac cones at the TRIM $(k_x,k_y)\in\{0,\pi\}^2$.
Applying our recursive spectral localizer approach, we choose $G_1 = X$ and $G_2 = Y$ to confine in real space, and define $G_3(\vec k)=3-2(\cos k_x+\cos k_y)$ to isolate the Dirac point at $\Gamma = (0,0)$. $G_3(\vec k)$ is $-1$ at $\Gamma$ and is positive at each of the other three Dirac points [Fig.~\ref{fig:localizersemimetal}(e--h)].

Interestingly, the model Eq.~\eqref{eq:minimaldirac} is characterized by two topological invariants. TRS alone puts the model in class AII. The resulting chain of classes is $\mathrm{AII}\rightarrow\mathrm{DIII}\rightarrow\mathrm{D}\rightarrow\mathrm{BDI}$ (from Table~\ref{tab:symmetry_clock}), and the $\bbZ_2$ index is the determinant sign of Eq.~\eqref{eq:detinvariant}. When chiral symmetry, $C=\sigma_z$, is also taken into account, the model belongs to class DIII, and the $\mathbb{Z}$ index is given by the half-signature of Eq.~\eqref{eq:sigindex} since the corresponding zero-dimensional Hamiltonian belongs to class AI. Notably, this second invariant does not require TRS.

We test each index by breaking one symmetry at a time, either $C$ or $T$. $-\lambda\sigma_x$ breaks $T$ (but preserves $C$) and shifts the Dirac cone away from $\Gamma$. In contrast, the constant energy shift $\lambda\,\id$ breaks only $C$, transforming the Dirac cone at $\Gamma$ into a circle of zero-energy states whose radius increases with $\lambda$. In both cases, for sufficiently large $\lambda$, the zero-energy set moves outside the circle defined by $G_3$. The corresponding index then abruptly drops to zero, precisely when the localizer gap closes, as illustrated in Fig.~\ref{fig:localizersemimetal}(i,j).

\section{Mixed position-momentum localizers: valley Hall topology}
\label{sec:valley}

\subsection{Valley topology in graphene}

So far, we have considered spectral localizers that are diagonal in either position space or momentum space. However, the method readily extends to localizers that are diagonal in neither basis. It is particularly well suited to models in which an isolated mode in momentum space undergoes a topological transition across a domain wall. A typical example is provided by the quantum valley Hall effect with the observation~\cite{longju2015,huangke2024} of chiral edge modes between AB and BA regions in bilayer graphene. We use below an even simpler model of graphene with a staggered mass, i.e.\ opposite on-site energies on the two sublattices, that varies smoothly across a domain wall along $x=0$,
\begin{equation}
\begin{split}
H={}&m(X)\otimes\sigma_z
-t\bigl(\id+T_x^\dagger+T_y^\dagger\bigr)\otimes\sigma_+
\\[2mm]
&-t\bigl(\id+T_x+T_y\bigr)\otimes\sigma_- ,
\\[2mm]
&m(X)=m_0\tanh\left(\frac{X}{L_0}\right)+\lambda,
\label{eq:2Dvalley}
\end{split}
\end{equation}
where we use the same notation as in Eq.~\eqref{eq:semimetaltopo}.
The mass term opens a gap $2|m|$ in both valleys $K$ and $K'$ of graphene.
Near each valley, $H$ is a Dirac Hamiltonian whose mass $m(x)$ changes sign at the domain wall. The wall therefore binds a chiral channel in each valley that disperses along $y$ and crosses zero energy, the two-dimensional Jackiw--Rebbi bound state \cite{JackiwRebbi1976,Semenoff_2008,Martin_2008,Yao_2009,Zhang_2013}. The two valley edge modes propagate in opposite directions and are exchanged by time reversal.
The existence of zero-energy domain-wall modes requires a confining operator in position space, but this alone is insufficient to probe the topology as the two valleys carry opposite valley Chern numbers. A topological index that does not resolve momentum space thus yields a trivial result.

\begin{figure}[htbp]
    \centering
    \includegraphics[width=\linewidth]{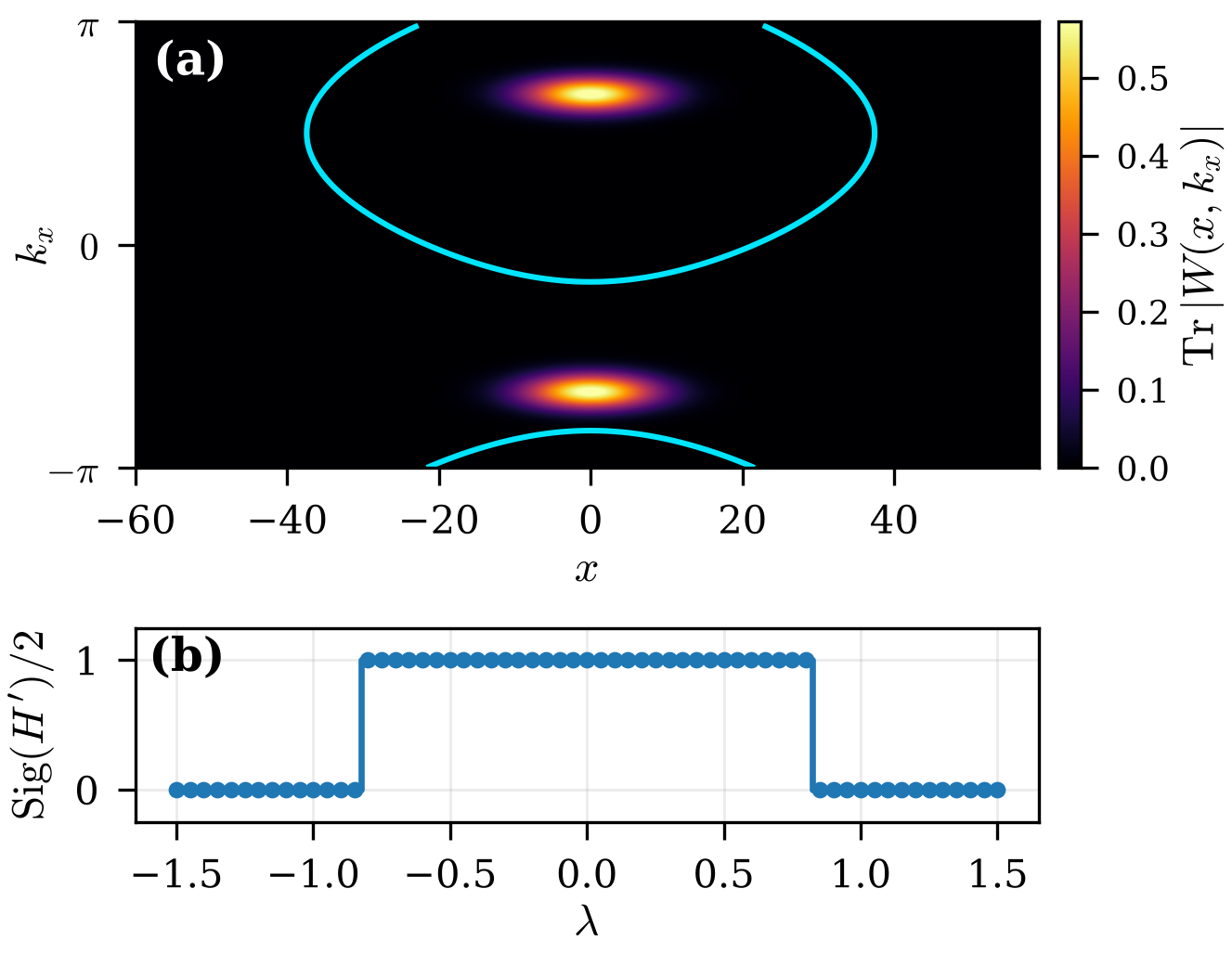}
    \caption{(a) $\mathrm{Tr}\,|W(x,k_x)|$ (Appendix~\ref{ap:wigner}) for the valley-Hall graphene ribbon \eqref{eq:2Dvalley}, showing the two wall modes at $k_x=\pm2\pi/3$; the cyan curve is the shell $g_2(x,k_x)=0$, which encloses one of them only. (b) Half-signature of the corresponding localizer, which changes at $|\lambda|\simeq0.8$ where the displaced wall crosses the shell; the same values are obtained on a $30\times12$ ribbon. Parameters: $L_0=L_x=120$ for (a), and a $30\times6$ ribbon ($L_0=L_x=30$) for the localizer; $m_0=3$, $t=1$, $\epsilon=0.14$, $x_0=0.18\,L_x$, $\kappa=0.4$.}
    \label{fig:LDOS2Dvalley}
\end{figure}

We therefore use two confining operators:
\begin{equation}
G_1=Y, \qquad
G_2=\left(\frac{X}{x_0}\right)^2+\ii(T_x^\dagger-T_x)-1 .
\end{equation}
to isolate both the domain wall in position space and the $K$ valley in momentum space. $G_1$ cuts the channel along $y$. The semiclassical symbol of $G_2$ is $g_2(x,k_x)=(x/x_0)^2-2\sin k_x-1$. At $x=0$, it is negative at the K point and positive at $K'$, and increases with $|x|$ such that its zeros form a closed path enclosing $K$. The model Eq.~\eqref{eq:2Dvalley} is represented by its phase-space density together with the lines of zeros of the semiclassical symbol $g_2$ in Fig.~\ref{fig:LDOS2Dvalley}. Since the wall modes extend along $y$, this density is drawn in the $(x,k_x)$ plane, after $y$ and $k_y$ have been eliminated by a partial trace over $y$ (see Appendix~\ref{ap:wigner} for details). Since $X$ and $T_x$ do not commute, $G_2$ is diagonal neither in position nor in momentum.
After the dimensional reduction, the spectral localizer takes the form
\begin{equation}
H'=H\otimes\eta_z+\kappa\, Y\otimes\eta_y+\kappa\, G_2\otimes\eta_x ,
\end{equation}
where the Pauli matrices $\eta_i$ enlarge the Hilbert space. Although time-reversal symmetry is preserved by the model Eq.~\eqref{eq:2Dvalley}, it is broken at the level of the individual valley $K$. The chain of symmetry classes is thus $\mathrm{A}\rightarrow\mathrm{AIII}\rightarrow\mathrm{A}$ and the topological index is the half-signature of $H'$ given in Eq.~\eqref{eq:sigindex}.
Its value is shown in Fig.~\ref{fig:LDOS2Dvalley}(b) as a function of a uniform shift $\lambda$ added to the staggered mass. This shift moves the domain wall to the position where $m_0\tanh(x/L_0)+\lambda=0$. As $\lambda$ increases, the domain wall eventually moves outside the confining shell defined by $G_2$, as reflected by the drop in the topological index.

Importantly, valley topology and the resulting chiral edge modes are not robust against arbitrary disorder but rely instead on a sufficiently smooth interface to suppress direct scattering between the two valleys~\cite{Martin_2008,Zhang_2013,longju2015,huangke2024}. For the localizer, the valley index is defined as long as the localizer gap stays open. Disorder or a sharp interface couples the two valleys through Fourier components of the potential at the wavevector $K-K'$ that separates them. When this coupling becomes comparable to the localizer gap, the gap closes and the index is no longer defined. The same mechanism splits the pair of zero modes of the Kitaev chain discussed below.

\subsection{Zero-hopping Kitaev chain}

We consider a second example with a Kitaev chain~\cite{Kitaev_2001} in the unconventional limit of zero hopping, with a spatially varying chemical potential. The Hamiltonian is
\begin{equation}\label{eq:kitaev}
    H=\sum_j m(x_j)\Bigl(c^\dagger_j c_j-\tfrac12\Bigr)
    -\frac{\Delta}{2}\sum_j\bigl(c^\dagger_{j+1}c^\dagger_j+c_j c_{j+1}\bigr),
\end{equation}
corresponding to the Bogoliubov--de Gennes form
\begin{equation}
  \begin{split}
    H={}&m(X)\otimes\tau_z
+\Delta\frac{T-T^\dagger}{2\ii}\otimes\tau_y,
\\[1mm]
&m(X)=m_0\tanh\left(\frac{X}{L_0}\right)+\lambda ,
  \end{split}
\end{equation}
 where $\tau_i$ act in particle-hole Nambu space $(c_j,c^\dagger_j)$ and a constant energy shift $\lambda$ has been added to the mass. %
For a homogeneous mass term $m$, the spectrum is given by
$E^2=m^2+\Delta^2\sin^2 k$. The gap is therefore $2|m|$ at the two time-reversal-invariant momenta, $k=0$ and $k=\pi$, where the pairing term vanishes. These two momenta  play the role of valleys. Near them, the Hamiltonian $H$ takes the form of a  one-dimensional Dirac model $m(x)\tau_z\pm\Delta \, q\,\tau_y$, where $q$ denotes the momentum measured relative to the corresponding valley $\pm$. The mass term changes sign at $x=0$ (for $\lambda=0$) and therefore binds two Jackiw--Rebbi zero modes~\cite{JackiwRebbi1976}, one per valley. Their localization length is of order $\sqrt{\Delta L_0/m_0}$.%

\begin{figure}[htbp]
    \centering
    \includegraphics[width=\linewidth]{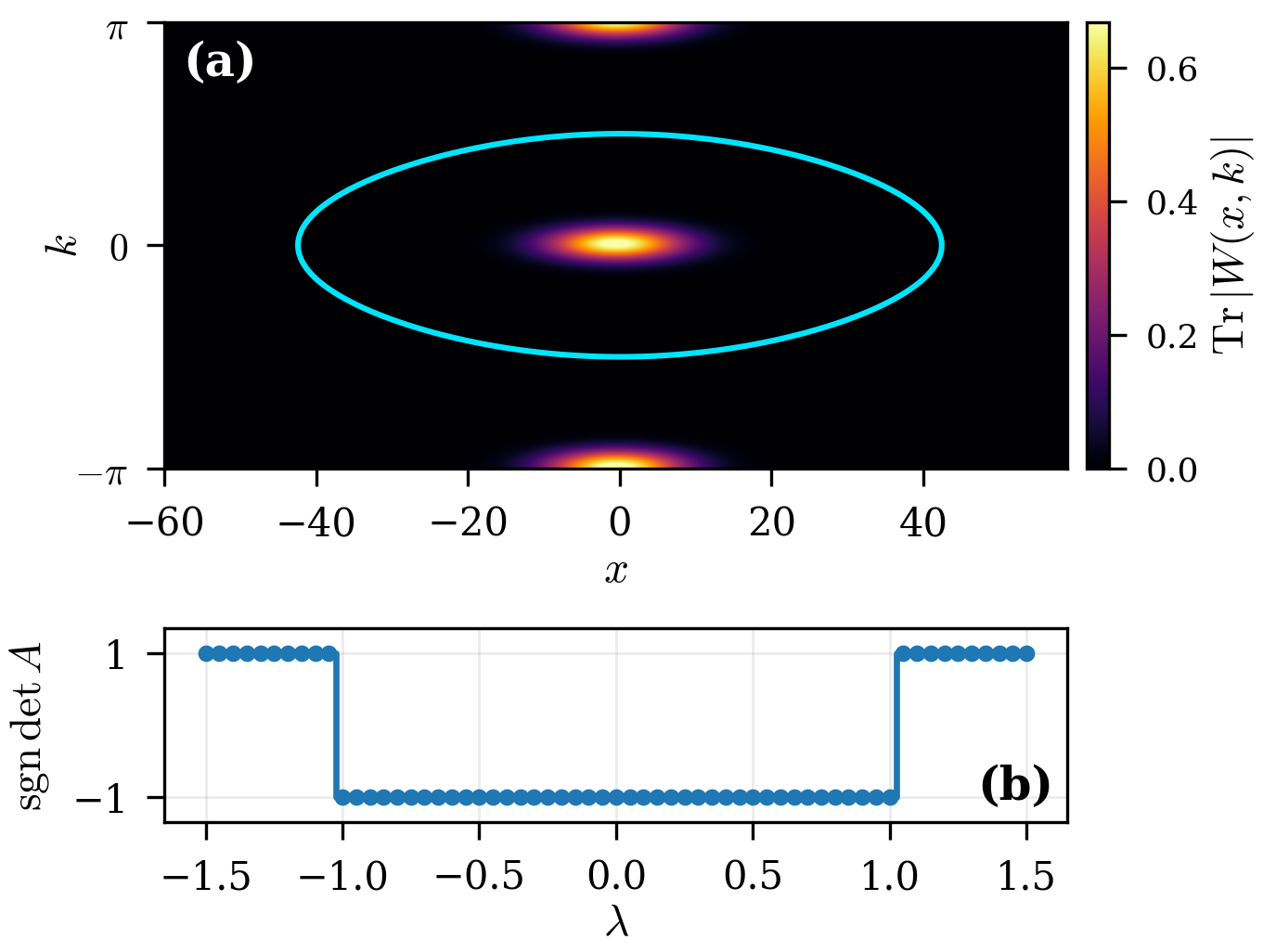}
    \caption{(a) $\mathrm{Tr}\,|W(x,k)|$ (Appendix~\ref{ap:wigner}) for the Kitaev chain \eqref{eq:kitaev}; the cyan curve is the shell $g(x,k)=0$, which encloses the mode near $(x,k)=(0,0)$ and excludes the one at $k=\pm\pi$. (b) Determinant-sign invariant of the corresponding localizer, which changes at $|\lambda|\simeq1.02$, where the displaced wall crosses the shell. Parameters: $L_0=L=120$, $m_0=3$, $\Delta=1$, $\epsilon=0.085$, $x_0=0.25\,L$, $\kappa=0.4$.}
    \label{fig:valleykitaev}
\end{figure}

The zero-hopping limit implies that the model Eq.~\eqref{eq:kitaev} is never in a topological phase. Its winding number integrates to zero as the two valleys contribute with opposite values. Instead, a momentum-selective index reveals a non-trivial topology. We confine the model with the single operator
\begin{equation}
G=\left(\frac{X}{x_0}\right)^2-(T+T^\dagger).
\end{equation}
Its semiclassical symbol $g(x,k)=(x/x_0)^2-2\cos k$ is negative at $(0,0)$ and positive at $(0,\pi)$ and at large $|x|$. The zeros of $g(x,k)$ enclose $(x,k)=(0,0)$, both the domain wall at $x=0$ and the valley around $k=0$. This is illustrated in Fig.~\ref{fig:valleykitaev}(a) with the phase-space density (Appendix~\ref{ap:wigner}) and the line of zeros.

The spectral localizer Hamiltonian is thus
\begin{equation}
H'=H\otimes\eta_z+\kappa\, G\otimes\eta_x.
\end{equation}
The Kitaev model and the selected valley are in class D~\cite{Sato_2017}, with the particle-hole symmetry operator $P=\tau_x\Kc$, $P^2=+\id$. After dimensional reduction, $\mathrm{D}\rightarrow\mathrm{BDI}$, $H'$ acquires the chiral symmetry with the operator $C=\eta_y$. In the Majorana basis where $P$ is the complex conjugation, $H$ is purely imaginary while $G$ is real. In the chiral basis of $H'$, the off-diagonal block can be chosen, up to an overall phase, as the real matrix
\begin{equation}
    A=\ii H+\kappa\, G ,
\end{equation}
and the zero-dimensional $\bbZ_2$ invariant is the determinant sign $\sgn\det A$ of Eq.~\eqref{eq:detinvariant}. Increasing the energy shift $\lambda$ moves the domain wall outside the circle of zeros. It is marked by a sudden change in the index corresponding to a gap closing of the spectral localizer Hamiltonian $H'$, as shown in Fig.~\ref{fig:valleykitaev}(b).

\begin{figure}[htbp]
    \centering
    \includegraphics[width=\linewidth]{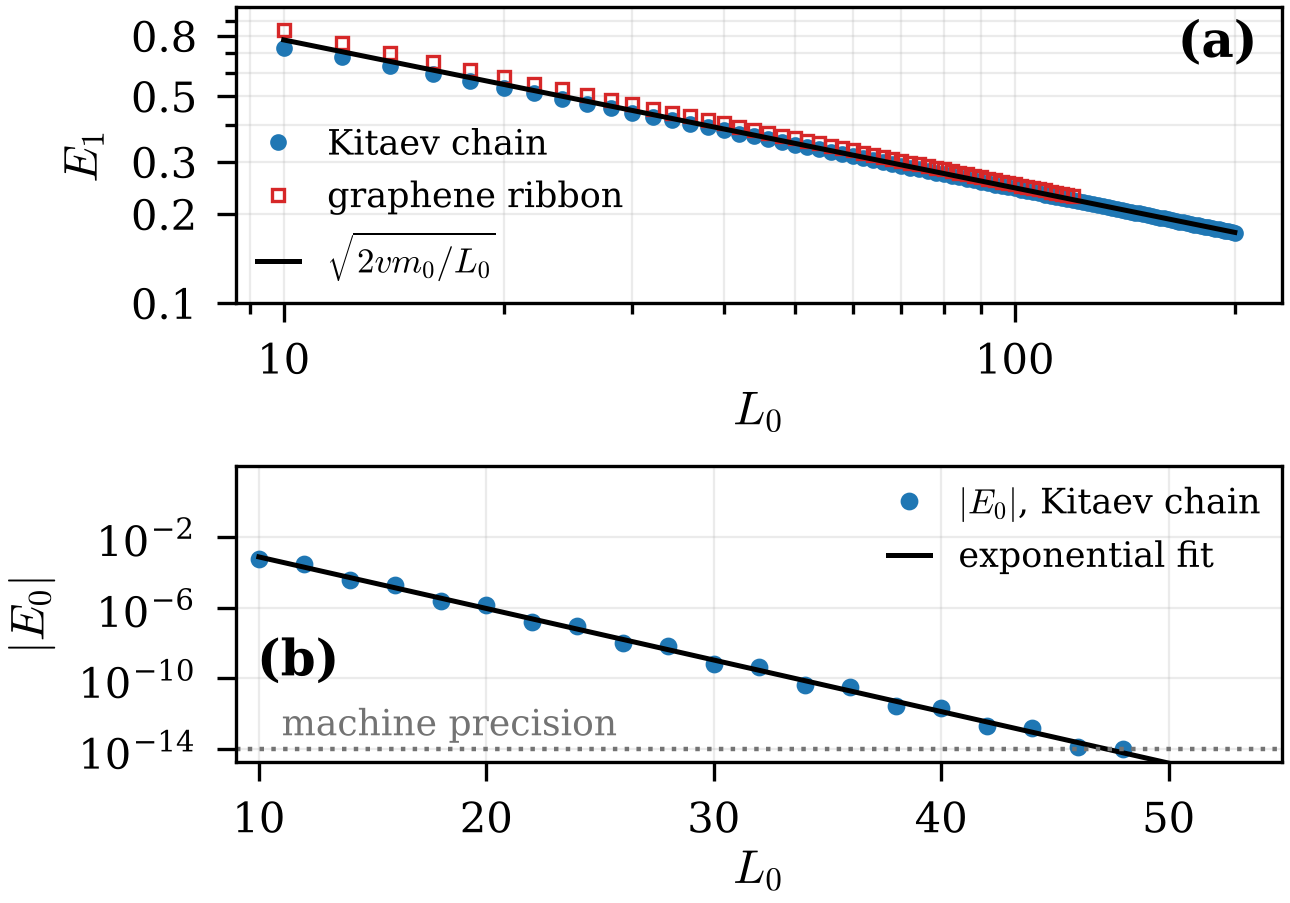}
    \caption{The two ways a Volkov--Pankratov level goes to zero as the wall is smoothed, at $\lambda=0$ with $m_0=3$ and $v=\Delta=t=1$. (a) The gap of the first massive rung follows $\sqrt{2v m_0/L_0}$ for both models. (b) The gap of the topological $n=0$ pair, measured on the Kitaev chain, is exponential in $L_0$, down to machine precision (dotted line). The solid line in (a) is the prediction $\sqrt{2vm_0/L_0}$, not a fit, and the one in (b) is an exponential fit.
 The wall width and the system size are varied together, $L_0=L$.}
    \label{fig:vp}
\end{figure}

The pair of Jackiw--Rebbi zero modes is protected by the valley-selective topological index, provided the domain wall is sufficiently smooth on the lattice scale. More precisely, the two valleys can only be coupled through the Fourier component of the mass profile $m(x)$ at the momentum transfer $\Delta k=\pi$. This coupling lifts the degeneracy of the zero-energy pair, but its amplitude is exponentially suppressed with increasing domain-wall width $L_0$.
In addition to the zero-energy modes, the smooth mass profile also gives rise to a ladder of excited states within the bulk gap. This can be understood by linearizing the mass term near the domain wall, $m(x)\simeq m'(0)x$. The resulting Dirac model is exactly solvable and yields the spectrum
\begin{equation}
    E_n=\sqrt{2n\, v\, m'(0)}\,, \qquad n=0,1,2,\dots
    \label{eq:vpladder}
\end{equation}
in each valley, where $v$ is the velocity of the Dirac model normal to the wall: $v=\Delta$ for the Kitaev chain and $v=t$ for the graphene ribbon, $x$ being counted in unit cells of the oblique lattice. The $n=0$ level corresponds to the Jackiw--Rebbi zero mode, while the massive levels with $n\geq 1$ are analogous to the Volkov--Pankratov states~\cite{VolkovPankratov1985,Tchoumakov_2017}. Since $m'(0)=m_0/L_0$, the gap to the first massive level decreases with increasing $L_0$, as $L_0^{-1/2}$. This decay is nevertheless much slower than the exponentially small splitting of the two zero modes. Figure~\ref{fig:vp} compares these scalings: the first massive level of both the graphene ribbon and the Kitaev chain follows $\sqrt{2vm_0/L_0}$, while the splitting of the zero modes, measured on the Kitaev chain, decreases exponentially. The first massive level is also the gap $E_{\mathrm{gap}}$ of Eq.~\eqref{eq:deltawindow} for these examples, so that smoother walls require a narrower filter, $\epsilon\ll\sqrt{2vm_0/L_0}$.

\section{Conclusion}

We have shown how the spectral localizer can be used beyond the usual setting of topological insulators. The construction can be understood as a dimensional reduction: starting from a Hamiltonian with topological gapless modes, one adds confining operators $G_j$ until the final Hamiltonian $H'$ is fully gapped. The topology of $H'$ is then a zero-dimensional invariant, such as a signature, the sign of a determinant, or the sign of a Pfaffian.

The important step is the choice of the operators $G_j$. For ordinary topological insulators, they are shifted position operators and the localizer probes a point in the bulk. For more general systems, their phase-space symbols can be chosen so that the common zero set forms a shell around the modes of interest. This gives a direct way to apply the localizer to higher-order topological insulators, topological semimetals, and systems exhibiting valley-selective topology.

The examples discussed in this work show that the same construction detects both $\bbZ$ and $\bbZ_2$ indices. In the higher-order case, the shell encloses hinge modes. In the semimetal case, it isolates a single Dirac cone, or a node of higher order, in momentum space. For valley-selective topology, it selects modes localized in both position and momentum, as for the zero modes of a Kitaev chain without hopping, whose global winding number vanishes while the valley-selective index does not. In each case, a change of the localizer index is tied to a closing of the localizer gap, which occurs when the shell no longer separates the chosen modes from the rest of the spectrum.

This formulation is useful because it does not require translation invariance. It can therefore be applied directly to finite systems and to systems with disorder or spatial inhomogeneity, as recently probed experimentally in higher-order topological lattices \cite{Dureth_2025}. It also keeps the computation at the level of sparse finite matrices, for which the signature and the Pfaffian can be obtained from a single $LDL^\dagger$-type factorization \cite{Cerjan_2024tutorial,Wimmer_2012}. The spectral localizer thus provides a practical diagnostic for topological modes that are localized in real space, momentum space, or both.

Two limitations of the construction are worth emphasizing. First, it depends on the choice of $\kappa$: although the criterion \eqref{eq:gapcriterion}  guarantees the existence of a gap, it is far from being necessary. In practice, one must therefore check that the index remains stable over a finite range of $\kappa$. Second, the phase-space shell must be specified by hand, which requires prior knowledge of where the modes of interest are localized in phase space. In this sense, the localizer certifies a proposed phase-space region rather than identifying it in the first place.
Natural extensions include a systematic treatment of strong or intervalley disorder, for which the phase-space shell would have to be defined in terms of an appropriately averaged symbol, and an extension to interacting systems, where the zero-dimensional invariants employed here have no straightforward counterpart.

\begin{acknowledgments}
We thank Alex Cerjan, Adam Yanis Chaou, Adolfo Grushin, Terry Loring, and Luca Trifunovic for helpful discussions. L.J. was supported by the European Research Council (ERC) under the European Union's Horizon 2020 research and innovation programme (Grant Agreement No.~101001902), by the Swedish Research Council (VR) through Grant No.~2020-00214, and by a travel grant from the Roland Gustafssons Stiftelse f\"{o}r teoretisk fysik.
\end{acknowledgments}

\appendix
\section{Topological clock of the localizer}
\label{ap:Bottclock}

The spectral-localizer construction $H \to H'$ induces a systematic shift in the Altland--Zirnbauer (AZ) symmetry class \cite{AltlandZirnbauer1997}. This mapping can be viewed as a clock on the periodic table of topological insulators \cite{Kitaev2009,Ryu2010,TeoKane_2010,Chiu_2016}. For defects localized in position, enclosed by a sphere in position space, it reproduces the classification of Teo and Kane~\cite{TeoKane_2010}, as discussed in Sec.~\ref{sec:teokane}. Each confining operator lowers $\delta$ by one and shifts the symmetry class in a predictable way when the added operator $G$ is real, as for $X_j$ and for the operators of Secs.~\ref{sec:hoti} and~\ref{sec:valley}. A complex $G$, such as $G_3$ in graphene, breaks the anti-unitary symmetries, and the reduction then starts from class A or AIII.

Without anti-unitary symmetry, the two maps \eqref{eq:7a} and \eqref{eq:7b} simply alternate between classes A and AIII: $H\otimes\sigma_z+\kappa\,G\otimes\sigma_y$ anticommutes with $\id\otimes\sigma_x$, while $H+\kappa\,CG$ has no symmetry left. With anti-unitary symmetries, the construction of $H'$ follows two paths, depending on the symmetries of $H$. If the system has both time-reversal symmetry $T$ and particle-hole symmetry $P$, we define the chiral operator $C=TP$ and assume $C^2=\id$.

\paragraph*{Non-chiral case:}
If $H$ has only particle-hole symmetry, with $P^2=\pm\id$, the augmented Hamiltonian $H'=H\otimes\sigma_z+\kappa\,(G\otimes\sigma_y)$ acquires both a particle-hole symmetry $P'=P\otimes\id$ with $(P')^2=\pm\id$ and a time-reversal symmetry $T'=P\otimes\sigma_x$ with $(T')^2=P^2\otimes\sigma_x^2=\pm\id$.

Conversely, if $H$ has only time-reversal symmetry, with $T^2=\pm\id$, the same localizer has a time-reversal symmetry $T'=T\otimes\sigma_z$ with $(T')^2=T^2$ and a particle-hole symmetry $P'=T\otimes(\ii\sigma_y)$ with $(P')^2=T^2\otimes(-\id)=\mp\id$. In both cases, the localizer adds the missing anti-unitary sector and moves the system into a chiral or fully symmetric AZ class.

\paragraph*{Chiral case:}
If $H$ has both $T$ and $P$, the localizer $H'=H+\kappa\,CG$ preserves only one anti-unitary sector. The outcome follows from $C^2=TPTP=\id$, which fixes the commutation algebra.

When $T^2=P^2$, the condition $T^2P^2=1$ implies $T[P,T]P=0$, hence $[T,P]=[T,C]=0$. In this case, $H'$ preserves the time-reversal symmetry $T'=T\otimes\id$, while the $CG$ term breaks $P$.

When $T^2=-P^2$, the condition $T^2P^2=-1$ implies $T\{P,T\}P=0$, hence $\{T,P\}=\{T,C\}=0$. In this case, $H'$ preserves the particle-hole symmetry $P'=P\otimes\id$ and breaks $T$.

These class shifts are consistent with Bott periodicity at the level of symmetry classes, as in the Clifford-extension argument of Ref.~\cite{Kitaev2009}. By shifting the class, the localizer transforms the $\bbZ$ or $\bbZ_2$ invariant of the gapless modes into a zero-dimensional invariant of the gapped operator $H'$. The corresponding class transitions are summarized in Table~\ref{tab:symmetry_clock}.

\section{Phase-space densities}
\label{ap:wigner}

Every phase-space picture in this work is built from the Gaussian spectral filter
\begin{equation}
    \rho_\epsilon = \mathrm{e}^{-H^2/\epsilon^2},
    \label{eq:gaussfilter}
\end{equation}
rather than from a projector assembled out of a few selected eigenstates. The reason is locality: \eqref{eq:gaussfilter} is a smooth function of $H$, so for a local Hamiltonian its kernel $\rho_\epsilon(x,x')$ decays rapidly in $|x-x'|$, whereas the kernel of a sharp energy window may be long range. This avoids a known artifact of phase-space quasi-probability distributions such as the Wigner function: two modes far apart in phase space produce interference fringes halfway between them, where there is no low-energy state. A kernel that decays fast suppresses these fringes \cite{loring2021locality}, so that $W(x,k)$ shows where the modes sit in phase space.

The width $\epsilon$ must be large enough to keep the whole set of modes enclosed by the shell and small enough to reject the rest of the spectrum,
\begin{equation}
    |E_0| \;\ll\; \epsilon \;\ll\; E_{\mathrm{gap}} ,
    \label{eq:deltawindow}
\end{equation}
where $E_{\mathrm{gap}}$ is the gap of $H$ on the shell and $E_0$ the residual finite-size energy of the enclosed modes, which vanishes in the thermodynamic limit. The value used is quoted in each caption.

Labeling the lattice sites by $x$ and the internal degrees of freedom (sublattice, orbital, spin or Nambu index) by $\alpha$, the kernel of \eqref{eq:gaussfilter} is a matrix in these internal indices, $\rho_{\alpha\beta}(x,x')=\langle x,\alpha|\rho_\epsilon|x',\beta\rangle$, normalized to unit trace. We use its one-sided transform, which keeps this matrix structure,
\begin{equation}
    W_{\alpha\beta}(x,k) = \sum_r \rho_{\alpha\beta}(x+r,\,x)\, \mathrm{e}^{-\ii k r} ,
    \label{eq:wignerdef}
\end{equation}
so that at each point of phase space $W(x,k)$ is a matrix acting on the internal degrees of freedom. Up to normalization, $W(x,k)=N\langle k|\rho_\epsilon|x\rangle\langle x|k\rangle$ is the Kirkwood--Rihaczek (or Kirkwood--Dirac) quasi-probability distribution~\cite{Kirkwood_1933,Rihaczek_1968}, a variant of the Wigner function~\cite{Case_2008} with the same marginals. Its marginals are $\sum_k W(x,k)/N=\rho(x,x)$, the density matrix at site $x$, and $\sum_x W(x,k)$, proportional to the occupation at wavenumber $k$, both as matrices in $\alpha$. We use this one-sided convention rather than the symmetric one of the Wigner function, $\rho_{\alpha\beta}(x+r/2,x-r/2)$: it keeps $x$ on lattice sites instead of introducing artificial midpoints, and it carries over to two or more dimensions without further bookkeeping. The price is that $W(x,k)$ is not Hermitian, and the figures show its trace norm,
\begin{equation}
    \mathrm{Tr}\,|W(x,k)| = \mathrm{Tr}\sqrt{W(x,k)\,W(x,k)^\dagger} ,
    \label{eq:wignernorm}
\end{equation}
the sum of its singular values, which is positive and does not depend on the choice of internal basis. For the wall modes of Figs.~\ref{fig:LDOS2Dvalley} and~\ref{fig:valleykitaev}, each of which has a fixed internal structure, it coincides to within $10^{-3}$ with $|\sum_\alpha W_{\alpha\alpha}(x,k)|$.

In $d$ dimensions, $W(\vec r,\vec k)$ depends on $2d$ variables, and every two-dimensional picture of this work is a marginal of $W$: the dimensions that are not drawn are summed over. Summing over the momenta $(k_x,k_y)$ gives the local density, $\sum_{\vec k}W(\vec r,\vec k)/N=\rho(\vec r,\vec r)$, drawn in the $(x,y)$ plane in Figs.~\ref{fig:dimensional-reduction} and~\ref{fig:localizersemimetal}(a,c,e,g), and in three dimensions in Fig.~\ref{fig:HOTI}(a,b). Summing over the positions $(x,y)$ gives the momentum density, drawn in Fig.~\ref{fig:localizersemimetal}(b,d,f,h). Summing over $y$ and $k_y$ gives a picture in the $(x,k_x)$ plane, as in Fig.~\ref{fig:LDOS2Dvalley}(a). In practice these marginals are computed directly from the kernel. Since $\sum_{k_j}\mathrm{e}^{-\ii k_j r_j}=N_j\,\delta_{r_j,0}$, summing over a momentum $k_j$ keeps only the diagonal of the kernel in $x_j$, and summing over a pair $(x_j,k_j)$ amounts to the partial trace of the kernel over $x_j$. Traced over the internal indices, the local and momentum densities are thus $n(\vec r)=\sum_\alpha\rho_{\alpha\alpha}(\vec r,\vec r)$ and $n(\vec k)=\sum_{\alpha}\sum_{\vec r,\vec r\,'}\mathrm{e}^{-\ii \vec k\cdot(\vec r-\vec r\,')}\rho_{\alpha\alpha}(\vec r,\vec r\,')$, the latter being defined whether or not the operator is translation invariant, and the picture in the $(x,k_x)$ plane is
\begin{equation}
    \frac{1}{N_y}\sum_{y,k_y}W(x,y,k_x,k_y)=\sum_r\rho_x(x+r,x)\,\mathrm{e}^{-\ii k_x r},
    \label{eq:reducedwigner}
\end{equation}
where $\rho_x(x,x')=\sum_y\rho(x,y;x',y)$ is the kernel reduced over $y$. This is the one-dimensional transform \eqref{eq:wignerdef} of $\rho_x$, which is still a matrix in the internal indices and whose trace norm \eqref{eq:wignernorm} is shown. For Fig.~\ref{fig:LDOS2Dvalley}(a), the ribbon is taken periodic in $y$, so that $\rho_x=\sum_{k_y}\rho_{k_y}$ is the sum of the kernels of the Bloch Hamiltonians $H(k_y)$, evaluated at $181$ values of $k_y$; the localizer itself is computed with open boundary conditions in both directions.

The pictures of the partial localizers $H_m$ in Figs.~\ref{fig:dimensional-reduction}, \ref{fig:HOTI} and~\ref{fig:localizersemimetal} are obtained in the same way, with $H$ replaced by $H_m$ in the filter \eqref{eq:gaussfilter}. The auxiliary indices added by the Clifford matrices are then traced out together with the internal ones, so that the densities of $H$ and of $H_m$ are drawn on the same lattice.

\section{Rhombohedral multilayer graphene}
\label{ap:multilayer}

Each layer $l=1,\dots,n$ is a copy of the graphene model \eqref{eq:semimetaltopo}, with sublattices $A_l$ and $B_l$. In rhombohedral stacking, $B_l$ lies on top of $A_{l+1}$, and the hopping $\gamma_1$ between them acts within a unit cell of the oblique lattice. With the deformation of Sec.~\ref{sec:semimetal} added to every layer, $H(\lambda)=H-\lambda\sum_l\sigma_x^{(l)}$, the Bloch Hamiltonian is block tridiagonal in the layer index, in the basis $(A_1,B_1,\dots,A_n,B_n)$,
\begin{equation}
    H_{\vec k}(\lambda)=\begin{pmatrix}
    h_{\vec k}(\lambda) & \Gamma & & \\
    \Gamma^\dagger & h_{\vec k}(\lambda) & \ddots & \\
     & \ddots & \ddots & \Gamma \\
     & & \Gamma^\dagger & h_{\vec k}(\lambda)
    \end{pmatrix},
    \label{eq:multilayer}
\end{equation}
with the $2\times2$ blocks
\begin{equation}
    h_{\vec k}(\lambda)=\begin{pmatrix} 0 & f(\vec k)-\lambda\\ f(\vec k)^*-\lambda & 0\end{pmatrix},\qquad
    \Gamma=\begin{pmatrix} 0 & 0\\ \gamma_1 & 0\end{pmatrix},
\end{equation}
where $f(\vec k)$ is given in Eq.~\eqref{eq:semimetaltopo}. The deformation thus strengthens the intracell bond of each layer, $t\to t+\lambda$, and replaces $f$ by $f-\lambda$. The interlayer hopping connects the two sublattices, so that the chiral symmetry $C=\mathrm{diag}(1,-1,\dots,1,-1)$ is preserved. In the sublattice basis $(A_1,\dots,A_n,B_1,\dots,B_n)$, the Hamiltonian is block off-diagonal, and its $n\times n$ block $F_{\vec k}(\lambda)$ has $f-\lambda$ on the diagonal and $\gamma_1$ just below it. Since $F$ is triangular, the bands touch at zero energy only where $f(\vec k)=\lambda$, that is, at the positions of the deformed monolayer cones, where they form the nodes of order $n$ of Refs.~\cite{Min_2008,Koshino_2009}.

The localizer of Fig.~\ref{fig:localizersemimetal}(i,j) is computed for this full $2n$-band model, with $G_1$, $G_2$ and $G_3$ acting in the same way on all layers. At finite $\kappa$, the index drops from $-n$ to $0$ in $n$ steps instead of one. The reason is that the confinement by $G_1$ and $G_2$ turns the cone of each layer into one zero mode of the partial localizer $H_1$ [Fig.~\ref{fig:localizersemimetal}(c,d)]. Without interlayer hopping, these $n$ modes would all leave the shell at the monolayer transition $\lambda_c$. The hopping $\gamma_1$ couples them like the $n$ sites of an open chain and splits them into $n$ levels, $\gamma_1\cos[\pi j/(n+1)]$ with $j=1,\dots,n$. Each level acts as a shift of $\lambda$ for the corresponding mode, which therefore reaches the transition at $\lambda_j\simeq\lambda_c+\gamma_1\cos[\pi j/(n+1)]$. On the $15\times15$ lattice, this splitting is the same, to within $0.005\,t$, for $\kappa=0.05$, $0.1$ and $0.2$, since $\gamma_1$ stays small compared with the energy scale of the confinement, $\sqrt{\kappa t}$. For $\kappa\ll\gamma_1^2/t$, which requires larger lattices, the localizer instead resolves the node of order $n$ as a whole, and all the steps are expected to tend to the symbol prediction $\lambda=t$ as $\kappa\to0$.

\bibliography{biblio}

\end{document}